# Random Walks in Routing Landscapes

T.Michalareas.
Telecommunication System Research Group
Electrical & Electronics Department
University College London (UCL), London, UK

*Abstract*—**In this paper we present a combinatorial optimisation view on the routing problem for connectionless packet networks by using the metaphor of a landscape. We examine the main properties of the routing landscapes as we define them and how they can help us on the evaluation of the problem difficulty and the generation of effective algorithms. We also present the random walk statistical technique to evaluate the main properties of those landscapes and a number of examples to demonstrate the use of the method.**

## I. Introduction

The unicast routing problem is considered one of the challenging problems in the framework of QoS provisioning packet networks (see [1] for more on QoS). The problem is to compute routes for all defined source-destination pairs and all the supported services that satisfy two conditions: a) the routes' metrics satisfy the constraints (or optimisation targets set by the services definitions), b) the overall selection of routes, or so-called route configuration, achieves some global efficiency in resource usage. There are two main types of difficulty with solving this problem. The first type is associated with the type of constraints. For some selection of those constraints the problem becomes NP-complete (see [2], [3]), thus impossible to solve in an optimal way for large network instances. The second type of difficulty is due to the efficiency target of the routing problem and the nature of traffic statistics that flows such communication networks. The QoS routing algorithm is expected to operate online, to gather network state information (which cannot be complete) and respond to any statistical shifts (during traffic time-scales from minutes to days), that would change the optimal operational point for all types of services traffic that could possible have independent spatial and temporal distributions.

This complexity of finding the most efficient routing configuration necessitates the use of heuristics that could discover good-enough solutions to enforce before the network state has changed significantly (for an example of a heuristic see [4]). These heuristics should operate based on partial network state information and improve significantly the network resources utilization. In this paper we present a combinatorial view on the problem. We describe a technique based on the theory of landscapes (see [5]) that could be used to characterise the structure of this difficulty and help to select appropriate heuristics. At the first section of the paper we introduce the basic elements of landscape theory and how the routing problem can be fitted in this framework. At the rest of the paper we present the statistical technique of random walks, how it can be used to in landscapes as well as some examples for the routing problem based on a number of packet level simulations.

## II. Fitness Landscape Theory & Routing

The notion of a fitness landscape comes from biology, where it is used as a framework for thinking about evolution. A theory of landscapes is based on three basic components that need to be defined. There is a finite but large set *V* of possible solutions-configurations, which comprises the solution space of a problem. There is a neighbourhood relation that denotes which points of that space are each other neighbours and allows us to view the solution space *V* as the vertex set of a graph *G*. We will refer to *G* as the configuration space. The third component of the landscape is a "fitness function" $f : V \leftarrow R$, that defines a partial ordering relation $>_f$ on the solution space *V*.

An example of a fitness landscape can be defined for the Travelling Salesman Problem (TSP). In TSP a salesman starts from his home city and visits exactly once each of the n cities of a given list, then returns home. The solution space *V* is the set of the possible tours, i.e., all permutations of the cities on the salesman's list. The "fitness" $f$ of a particular tour $t$, is its total length. The neighbourhood relation is defined as follows. Two tours are neighbours if they can be inter-converted by a simple operation on the list of cities. Simple operations are considered swapping two cities in the permutation or inverting the order of a contiguous part of the list.

In order to define a fitness landscape for the routing problem we have to define these 3 components. The easiest part is the definition of the solution space $V$. Let $G$ be the graph corresponding to the topology of the network with $n$ nodes. Let $s$ be any source node and $d$ any destination node. We define $R(s, d)$ to be the set of alternative routes between these two nodes and $K_i$ to be the correspondent number of alternative routes. There are totally $N = n * (n - 1)$ possible s-d pairs. A point in the solution space is a vector of $N$ entries. Every entry holds the used route for the corresponding $s - d$ pair. In this case the size of the configuration space is bounded by the number of possible permutations of the $n$ nodes.

There is a number of ways to associate two route configurations in order to define the neighbourhood relation for the same problem and define the configuration space $G$. This neighbouring relation is basically an extension of the Hamming distance for $0-1$ *n-bit* strings to *n-state* strings with a number of $M_i$ alternative values for every entry $i$ of the string. Two neighbouring routing configuration have distance of 1. Generally two routing configurations have as a distance the number of different routes. This relation can be though as a first measure of the cost to transform a given configuration to another. The size of this configuration space is the product of the number of alternative routes for each $s - d$ pair over all $s - d$ pairs.

The fitness function $f$ depends on the version of the routing problem we are interested in. In our case we are interested in routes that minimize packet delay metric $f_1$. In the rest of the paper we will make our analysis based on this partial ordering. We are interested in two types of points and their distribution characteristics in the configuration space. There are the global optima, which are the points that have greater fitness values that any other point. There are also local optima that have the greatest value of fitness value in a neighbourhood of a predetermined distance $k$ around them. As basins we define the set of points between two local optima.

Any routing algorithm can be though as a search process in this fitness landscape. A brute-force algorithm would enumerate all the configuration space points, register their fitness values and select the best one. In versions of the routing problems where there are no shortcuts to be utilized by polynomial algorithms, to find an efficient solution without enumerating all of them the algorithm utilises a rule of thumb that can be used to locate local optima that have similar fitness values to the global optima. The characteristics of the distribution of local optima along with some other properties of the landscapes are important for determining good heuristics for the problem. As an example consider a fitness configuration landscape that has local optima clustered together in an average distance of k. In such a landscape once a local optimum is found the probability of finding another local optima by examining configurations that have distance greater than *k* decreases.

III. LANDSCAPE PROPERTIES AND THE RANDOM WALK TECHNIQUE

In this paper we are interesting in two aspects of the landscapes characteristics. The first one is the fitness distance correlation coefficients (FDC) and the second one is the autocorrelation function.

The fitness distance correlation coefficient (FDC) $r_{FDC}$, has been proposed as a summary statistic for measuring a problem's difficulty for Genetic Algorithms (see [6]). The definition of $r$ is given in Equation-1 for a set $F = f_1, f_2, ..f_n$ of measured fitnesses and a corresponding set of $D = d_1, d_2, ..d_n$ of the Hamming distances of the corresponding configurations to the best of them.

$$r_{FDC} = C_{FD}/(S_F * S_D). \qquad (1)$$

$S_F$ and $S_D$ are the standard deviations of the $F$ and $D$ set and $C_{FD}$ is their covariance. Similar to other correlation measures FDC values close to *0* indicate a random landscape where essentially there are no limits on the distance a search process should examine for better configurations than the currently selected. Although the FDC measure is not always a reliable summary statistic for the difficulty of finding optima in a landscape (see [6] for some known outliers), we

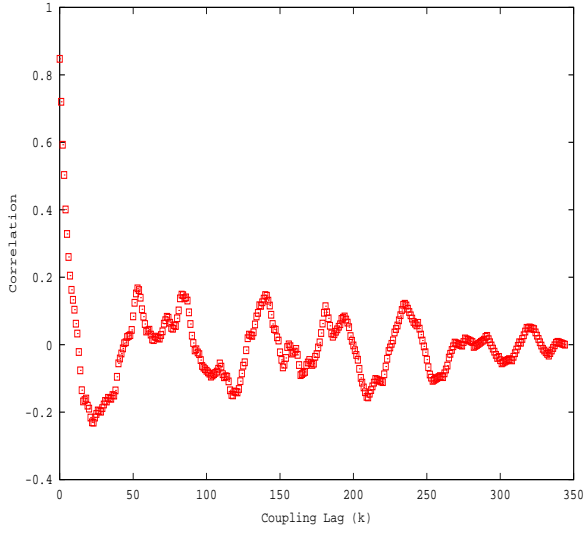 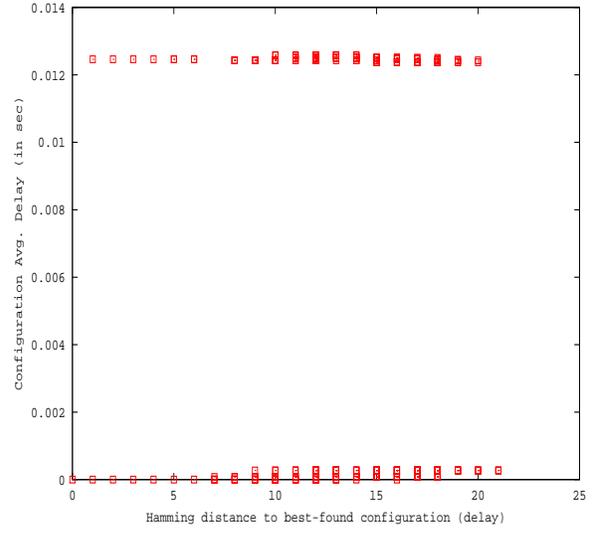

*a. Correlation Plot for Avg.Delay*     *b. Scatter Plot of Hamm. distance vs. Delay*

Fig. 1. HotSpot Traffic Scenario

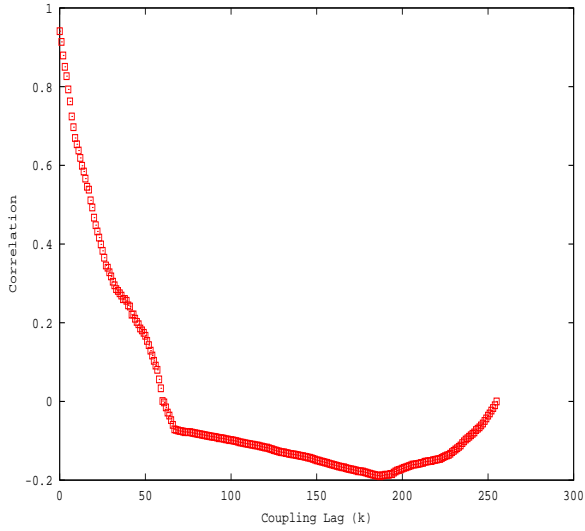 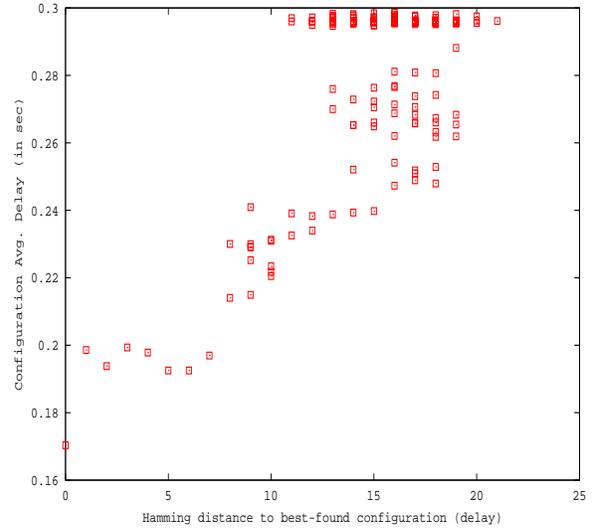

*a. Correlation Plot for Avg.Delay*     *b. Scatter Plot of Hamm. distance vs. Delay*

Fig. 2. Adjacent Nodes Traffic Scenario

will use it along with a scatter plot to understand the structure of the problem (for more details see examples).

The second measure we use is the autocorrelation function of the F set, $r(s)$ as in Equation-2.

$$r(s) = \frac{<f(x_t)f(x_{t+s})>_{x0,t} - <f(x_t)>^2_{x0,t}}{<f(x_t)^2>_{x0,t} - <f(x_t)>^2_{x0,t}} \quad (2)$$

The notation $<.>_{x_0,t}$ is used to denote the expectation taken over the set F (for all valid instances produces in time $t$) with first configuration $x_0$.

In order to estimate these properties of a landscape we use the random walk technique to generate the set $F$ as suggested in [7]. The random walk technique produces a number of routing configurations that we could evaluate their fitness. In the initial phase of the random walk a random routing configuration is selected. During the selection phase the next routing configuration is produced from the current one. We select an s-d pair in a uniform random way and its route is changed to one the rest of $K_i - 1$ alternative routes. The same step is repeated for n number of times, the length of the tour.

There are three characteristic shapes for the autocorrelation function. In the first one most of the function values are close to zero. This is a characteristic of a random landscape where there is no correlation between the fitnesses of the landscape. The second characteristic form of an autocorrelation function is the slow decaying form. In this case there is a weak correlation between the fitnesses. The third characteristic form is the fast decaying form that indicates a strong correlation among the fitnesses. Clearly problem instances with landscape autocorrelation of the third type are easier to be solved than that of the first and second type.

## IV. EXAMPLES

For the purpose of evaluating the random walk technique we have selected a simple network topology the cycle of a restrained number of nodes (6 nodes). In order to perform random walks we have selected to extend the ns-2 (see [2] for more) simulator version that supports MPLS. The routing walk technique has been implemented as an agent that gathers topology information, performs the computation of all the possible paths (enumeration), selects randomly (controlled by a random seed) an initial routing configuration and the in given time intervals selects randomly one viable source-destination pair and toggles its state to one of the other alternate routes.

We have selected to evaluate the properties of the landscape for the routing problem for two traffic patterns. The first traffic pattern formulates a logical star with one hot-spot node and the second has traffic only among adjacent nodes. In all scenarios our traffic is generated by CBR sessions established over UDP that produce 800 byte packets at 0.01 sec intervals. All the links have 1Mbit capacity and 10ms propagation delay. By using the FDC and the r(s) we will try to get some insight on the difficulty of optimising the routing configuration for this scenario.

Out topology although small (6 nodes) has 230 possible routing configurations. In order to evaluate the properties of the search space in addition to the autocorrelation function and FDC we use scatter plots. In the scatter plots we estimate for every random walk the best solution found for the delay metric. We then sorted the configurations according to an ascending order of the metric and estimated the Hamming distance of those configurations from the best found. At the x-axis of the scatter plot we have the hamming distance of the configurations and at the y-axis the value of the considered metric.

In Figure-1 we examine the case of the hot-spot traffic pattern. In Figure 1a we plot the autocorrelation function of the random walk collected samples for the delay metric. Clearly this is of the first type that indicates a random landscape with small correlation. The FDC statistic verifies that by giving as a value of -0.074350. In Figure 1b we have the scatter plot that shows clearly the structure of the problem. There are equally well solutions to the best one at almost all hamming distances from the best one found. That means there is no correlation structure on the landscape and a search process cannot use some bound on its search.

In Figure-2 we examine the case of the adjacent nodes traffic pattern. In Figure-2a we have the autocorrelation function of the landscapes that fits to the rapid decay pattern, which is indicative of strong correlation. The FDC statistic takes value of 0.67937, which indicates that the problem can be less difficult than that of the hot-pot scenario. In Figure-2b we have the corresponding scatter plot. There is clearly a large cluster formulated close to the best solution found bounded by Hamming distance of 7 units.

## V. CONCLUSIONS

We have presented a combinatorial view on the routing problem and the random walk technique to evaluate some of the properties of the resulting search space. Through a set of simple examples we have demonstrated how these properties can help us to understand the structure of the routing problem in search for better heuristics to solve routing problems.